\documentclass[aps,pre,showpacs,twocolumn]{revtex4}
\usepackage{bm}
\usepackage{graphicx}
\bibstyle{approve.bib}
\usepackage{amssymb}
\usepackage{epstopdf}
\usepackage{color}
\setcitestyle{super}
\usepackage{filecontents}

\newcommand{\be}{\begin{equation}}
\newcommand{\ee}{\end{equation}}
\newcommand{\bea}{\begin{eqnarray}}
\newcommand{\eea}{\end{eqnarray}}

\newcommand{\br}{\mathbf{r}}

\newcommand{\e}{\varepsilon}

\begin{document} 

\title{Controlling polymer translocation and ion transport via charge correlations}

\author{Sahin Buyukdagli$^{1}$\footnote{email:~\texttt{buyukdagli@fen.bilkent.edu.tr}} and T. Ala-Nissila$^{2,3}$\footnote{email:~\texttt{Tapio.Ala-Nissila@aalto.fi}}}
\affiliation{$^{1}$Institut de Recherche Interdisciplinaire USR3078 CNRS and Universit\'e Lille I, Parc de la Haute Borne, 52 Avenue de Halley, 59658 Villeneuve d'Ascq, France\\
$^{2}$Department of Applied Physics and COMP Center of Excellence, Aalto University School of Science, P.O. Box 11000, FI-00076 Aalto, Espoo, Finland\\
$^{3}$Department of Physics, Brown University, Providence, Box 1843, RI 02912-1843, U.S.A.}
\date{\today}

\begin{abstract}
We develop a correlation-corrected transport theory in order to predict ionic and polymer transport properties of membrane nanopores in physical conditions where mean-field electrostatics breaks down. The experimentally observed low KCl conductivity of open $\alpha$-Hemolysin pores is quantitatively explained by the presence of surface polarization effects. Upon the penetration of a DNA molecule into the pore, these polarization forces combined with the electroneutrality of DNA sets a lower boundary for the ionic current, explaining the weak salt dependence of blocked pore conductivities at dilute ion concentrations. The addition of multivalent counterions into the solution results in the reversal of the polymer charge and the direction of the electroosmotic flow. With trivalent spermidine or quadrivalent spermine molecules, the charge inversion is strong enough to stop the translocation of the polymer and to reverse its motion. This mechanism can be used efficiently in translocation experiments in order to improve the accuracy of DNA sequencing by minimizing the translocation velocity of the polymer.
\end{abstract}

\date{\today}
\maketitle

\section{Introduction}

The analysis of biopolymer sequences is of vital importance to understand the functioning of living organisms. Nanopore sensing methods that aim at sequencing biopolymers have drawn increasing attention during the last two decades. Since the seminal article from Kasianowicz et al.~\cite{e1},  electrophoretic polymer translocation through nanopores has been in central focus in this field. In addition to being a potentially fast and inexpensive method, this sequencing strategy based on the analysis of ionic current variations is also practical since it does not require the biochemical modification of the translocating DNA molecule. Intensive experimental work has been performed in order to refine the method of DNA sequencing with biological~\cite{e2,e3,e4,e5,e6} and solid-state nanopores~\cite{e7,e8,e9,e10,e11,e12}.  These works have revealed that ionic current variations and the accuracy of  their detection depend sensitively on the nanopore size and the translocation speed of DNA. Hence, it is of great importance to throughly characterize the optimal physical characteristics of the translocating polyelectrolyte and the pore in order to improve the resolution of this method.

The complexity of the polymer translocation problem stems from the entanglement between hydrodynamic DNA-solvent interactions, electrostatics correlations mediated by the confined electrolyte, and entropic costs associated with DNA conformations. Although hydrodynamic and entropic issues have been intensively addressed by previous Molecular Dynamics (MD) simulations~\cite{n1,n2,n3,n4,n5}, the central role played by electrostatics has not been scrutinized. Indeed, it is well established that the accurate sequencing of DNA necessitates a strong ionic current signal that also lasts long enough during the translocation event~\cite{e5}. The difficulty to realize these two combined conditions stems from the fact that a strong applied field resulting in a substantial signal will also cause the DNA molecule to escape too fast from the nanopore into the reservoir. In order to optimize the accuracy of this sequencing method, it is  thus crucial to figure out the physical conditions that minimize the velocity of DNA independently of the external voltage gradient. This requires in turn a consistent electrohydrodynamic theory of the charged liquid and the translocating polyelectrolyte. 

Previous MD simulation works ingenuously considered the effect of electrostatic correlations on the transport of ions~\cite{Qiao2005} and DNA molecules~\cite{Luan2009} thorough nanopores. In addition to their computational complexity, the canonical nature of these simulations fixing the total ion number in the nanopore does not correspond to the experimental setup where the ionic transport takes place via charge exchanges between the ion reservoir and the pore medium. The simplest approach that mimics polymer transport experiments consists in coupling the linear Poisson-Boltzmann (PB) theory with the Stokes equation. Within such a linear mean-field (MF) theory, electrostatic effects on the electrophoretic polymer translocation was investigated by \textit{Ghosal}~\cite{Ghosal2006,Ghosal2007}. In comparison with DNA translocation times measured for monovalent electrolytes~\cite{e9},  it was shown that translocation velocities are weakly affected by the monovalent salt density.  As revealed by additional MF studies~\cite{Keijan2009} and transport experiments~\cite{e12}, a more efficient way to control the DNA translocation velocity consists in tuning the pH of the solution. Nevertheless, this strategy is not universal since charge regulation effects depend sensitively on the chemical properties of the nanopore and the polymer type.  Within a solvent-explicit MD simulation approach, \textit{Luan} and \textit{Aksimentiev} showed that electrostatic correlations induced by strongly charged molecules can offer a more precise control over the DNA motion~\cite{Luan2009}. The correlation-induced mechanism is similar to the effect of multivalent charges on the decomplexation of oppositely charged polyelectrolytes~\cite{Antila2014}. However, the above-mentioned MF theories that neglect correlation effects are naturally unable to explore this possibility.

\textcolor{black}{Motivated by these facts, we develop the first unified theory of ionic and polymer transport beyond the PB approximation. First of all, our theory that can account for electrostatic many-body and surface polarization effects provides us with a realistic picture of ion transport through nanoscale pores where these effects are known to be non-negligible. Then, the ability of the proposed formalism in considering charge correlation effects allows to develop the idea of controlling polyelectrolytes via the charge reversal mechanism.} The computation of the ionic current through the nanopore  requires the knowledge of the charge partition between the polymer and the pore. Thus, we first extend the self-consistent (SC) electrostatic formalism previously developed for open pores~\cite{Buyuk2014} to the case of a cylindrical polyelectrolyte confined to the nanopore. By comparison with MC simulations, the SC scheme was shown to accurately handle electrostatic correlations including surface polarization effects~\cite{Buyuk2014}. Then, we couple the extended SC formalism with the Stokes equation. Within the framework of this correlation-corrected transport theory, we tackle three different questions. First, we investigate the physics behind the particularly low conductivity of open $\alpha$-Hemolysin pores observed in transport experiments~\cite{e3,Gu2000,Miles2001}. \textcolor{black}{We identify the underlying mechanism as the dielectric exclusion of ions, an effect inaccessible by previous MF theories.} Then, we scrutinize the non-monotonic salt dependence of DNA-blocked $\alpha$-Hemolysin pores~\cite{e3}. In the second part, we study the effect of charge correlations on the electrophoretic transport of polyelectrolytes and investigate the possibility to control the polymer translocation by tuning the reservoir ion densities in electrolyte mixtures with multivalent counterions. \textcolor{black}{We show that the addition of trivalent and quadrivalent ions into the solution allows to alter the speed and even the direction of translocating polyelectrolytes in a controlled way. We finally discuss the limitations and possible extensions of the present theory in the Conclusion.}

\section{Model and theory}
\label{theory}

\begin{figure}
\includegraphics[width=1\linewidth]{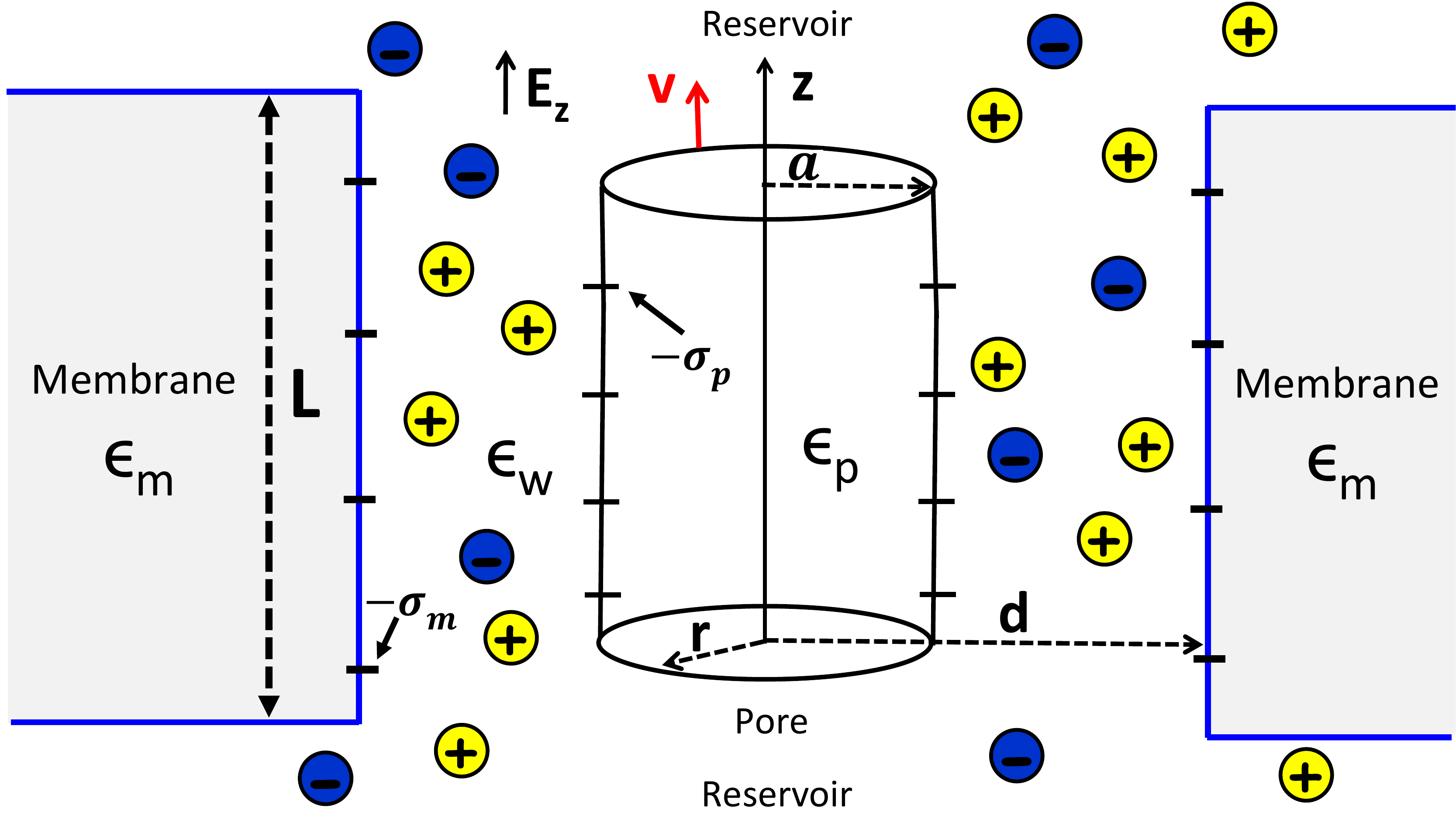}
\includegraphics[width=1\linewidth]{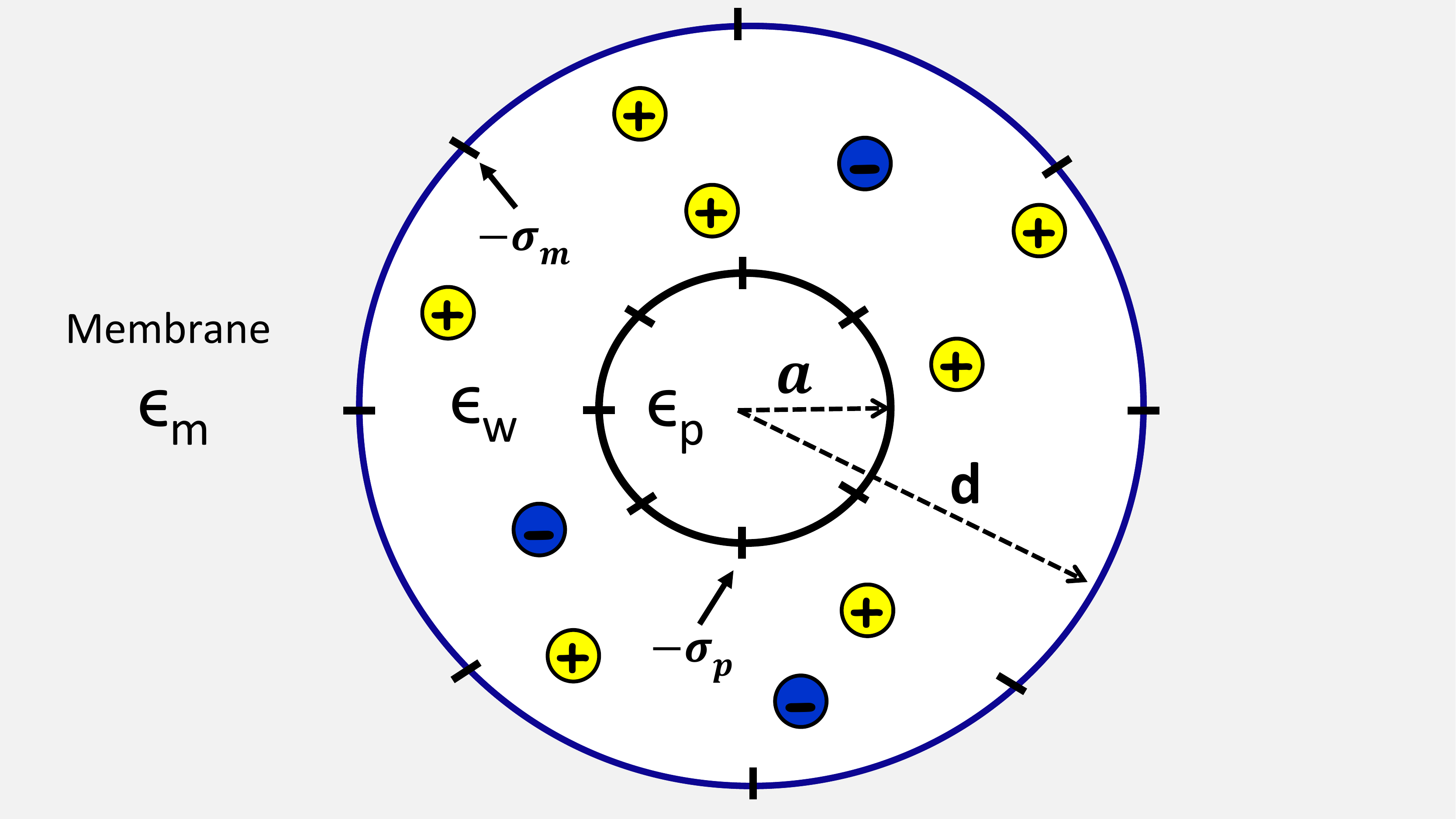}
\caption{(Color online) Schematic representation of the nanopore : side view (top plot) and top view (bottom plot). The cylindrical polyelectrolyte of radius $a$, surface charge $-\sigma_p$, and dielectric permittivity $\e_p$ is confined to the cylindrical pore of radius $d$, wall charge $-\sigma_m$, and membrane permittivity $\e_m$. Electrolyte permittivity is $\e_w=80$.}
\label{Fig1}
\end{figure}

\subsection{Translocating polymer model}

The geometry of the pore confining mobile charges and the polyelectrolyte is depicted in Fig.~\ref{Fig1}. The cylindrical nanopore of radius $d$ is in contact with an ion reservoir at the extremities. The polyelectrolyte is modelled as a rigid cylinder of radius $a$ whose longitudinal axis coincides with the axis of the nanopore. Both the nanopore and the polymer carry a smeared negative charge distribution with amplitudes $-\sigma_m$ and $-\sigma_p$, respectively. They also have corresponding static dielectric permittivities $\e_m$ and $\e_p$ that may differ from the permittivity of the electrolyte $\e_w=80$, all expressed in units of the air permittivity. The transport of the mobile charges and the polymer is driven by a potential gradient $\Delta V$ at the extremities of the channel. We assume that the potential decays linearly along the nanopore of length $L$, which results in a uniform electric field $E_z=\Delta V/L$ acting on the charges. We also emphasize that the nanopore lengths considered in the present work lie in the range $L\geq 10$ nm which is an order of magnitude larger than the Bjerrum length $\ell_B\simeq 7$ {\AA}. Thus, in the calculation of the ion densities, we will assume that the nanopore and the cylinder are infinitely long.  \textcolor{black}{We note that this approximation may not be accurate for membranes with nanoscale thickness comparable to the Bjerrum length. The consideration of charge transport effects through thin membranes is beyond the scope of this article.}

\subsection{Correlation-corrected ion densities}
\label{electro}

In this part we extend the SC theory of ion partition in open pores~\cite{Buyuk2014}  to the case of the cylindrical polyelectrolyte confined to the nanopore (see Fig.~\ref{Fig1}). Within this formalism, the total external potential induced by the pore and the polymer charges is composed of a modified PB contribution $\phi_0(\br)$ and a perturbative correction $\phi_1(\br)$ taking into account ionic cloud deformations, $\phi(\br)=\phi_0(\br)+l\phi_1(\br)$, where the loop expansion parameter $l$ will be set to unity at the end of the calculation. The components of the potential are obtained from the relations
\bea\label{eqt1}
&&\nabla\e(\br)\nabla\phi_0(\br)+\frac{e^2}{k_BT}\sum_{i=1}^pq_in_i(\br)=-\frac{e^2}{k_BT}\sigma(\br)\\
\label{eqt2}
&&\phi_1(\br)=-\frac{1}{2}\sum_iq_i^3\int\mathrm{d}\br'v_{el}(\br,\br')n_i(\br')\delta v_{el}^{(s)}(\br'),
\eea
which are coupled to the kernel equation for the electrostatic propagator accounting for charge fluctuations,
\bea\label{eqt3}
&&\nabla\e(\br)\nabla v_{el}(\br,\br')-\frac{e^2}{k_BT}\sum_{i=1}^pq_i^2n_i(\br)v_{el}(\br,\br')\nonumber\\
&&=-\frac{e^2}{k_BT}\delta(\br-\br').
\eea
In Eq.~(\ref{eqt1}),  the dielectric permittivity function is defined as $\e(\br)=\e_m\theta(r-d)+\e_w\theta(d-r)\theta(r-a)+\e_p\theta(a-r)$, $e$ is the elementary charge of the electron, $k_B$ the Boltzmann constant, $T$ the ambient temperature, and $q_i$ the valency of mobile ions with species $i$ ($i=1...p$ with $p$ the total number of ion species).  The auxiliary number density is defined as
\be\label{aux1}
n_i(\br)=\Omega(\br)\rho_{ib}e^{-q_i\phi_0(\br)}e^{-\frac{q_i^2}{2}\delta v_{el}^{(im)}(\br)},
\ee
where the function $\Omega(\br)=\theta(r-a)\theta(d-r)$ restricts the space accessible to the electrolyte, and $\rho_{ib}$ stands for the ionic concentration in the reservoir. Furthermore, in Eq.~(\ref{eqt1}), the fixed charge density is given by the function $\sigma(\br)=-\sigma_m\delta(r-d)-\sigma_p\delta(r-d)$. Then, the ionic self-energies in Eqs.~(\ref{eqt1})-(\ref{aux1}) are obtained from the renormalized equal-point Green's function
\be\label{self}
\delta v_{el}(\br)\equiv \lim_{\br'\to\br}\left\{v_{el}(\br,\br')-v_c^b(\br-\br')+\ell_B\kappa_b\right\},
\ee
with the Bjerrum length $\ell_B=e^2/(4\pi\e_w k_BT)$, the DH screening parameter $\kappa_b^2=4\pi\ell_B\sum_iq_i^2\rho_{ib}$, and the Coulomb potential in an ion-free bulk solvent $v_c^b(r)=\ell_B/r$. In terms of the total self-energy~(\ref{self}), the solvation and image-charge potentials in Eqs.~(\ref{eqt2})-(\ref{aux1}) are defined as
\bea
&&\delta v_{el}^{(s)}(\br)=\lim_{\e_{m,p}\to\e_w}\delta v_{el}(\br)\\
&&\delta v_{el}^{(im)}(\br)=\delta v_{el}(\br)-\delta v_{el}^{(s)}(\br),
\eea
where the image-charge contribution $\delta v_{el}^{(im)}(\br)$ includes surface polarization effects resulting from the dielectric discontinuities. From now on, we will exploit the cylindrical symmetry of the system where the electrostatic potentials depend exclusively on the radial distance $r$. In terms of the potentials introduced above, the local ion density in the pore is given by
\be\label{den1l}
\rho_i(r)=n_i(r)\left[1-lq_i\phi_1(r)-l\frac{q_i^2}{2} \delta v_{el}^{(s)}(r)\right],
\ee
which allows to derive the total charge density needed for the calculation of ionic currents 
\be
\rho_c(r)=\sum_iq_i\rho_i(r).
\ee
We finally note that the iterative solution scheme of Eqs.~(\ref{eqt1})-(\ref{eqt3}) introduced in Ref.~\cite{Buyuk2014} is unmodified by the presence of the polyelectrolyte.

\subsection{Pore conductivity and polymer velocity}
\label{trans}

Within the correlation-corrected electrostatic formulation introduced above, we now derive the polymer translocation velocity and the ionic conductance of the nanopore. The derivation of the polymer velocity will allow us to account for charge correlations and non-linearities neglected by the linear MF theory of Ref.~\cite{Ghosal2007}. The total velocity of an ion of species $i$ is composed of the convective and the drift velocities, $u_i(r)=u_c(r)+u_{Ti}(r)$. The drift velocity is given by
\be\label{ut}
u_{Ti}(r)=\mathrm{sign}(q_i)\mu_i\frac{\Delta V}{L},
\ee
where $\mu_i$ stands for the ionic mobility.  For comparison with ion transport experiments, we will take the experimentally established values $\mu_+=\mu_{K}=7.616\times10^{-8}$ $\mbox{m}^2/\mbox{V}^{-1}\mbox{s}^{-1}$ and $\mu_-=\mu_{Cl}=7.909\times10^{-8}$ $\mbox{m}^2\mbox{V}^{-1}\mbox{s}^{-1}$~\cite{book}.  The convective flow velocity $u_c(r)$ obeys in turn the Stokes equation
\be\label{s1}
\eta\Delta u_c(r)+e\rho_c(r)\frac{\Delta V}{L}=0,
\ee
with the viscosity coefficient of water $\eta=8.91\times 10^{-4}\;\mathrm{Pa}\;\mathrm{s}$. By acting with the Laplacian operator on Eq.~(\ref{eqt2}) and using the resulting equation with the relations~(\ref{eqt1})-(\ref{eqt3}),  one finds that the electrostatic potential is related to the charge density via the Poisson equation $r^{-1}\partial_rr\partial_r\phi(r)+4\pi\ell_B\rho_c(r)=0$.  Using this relation in Eq.~(\ref{s1}) to eliminate the charge density, the Stokes equation takes the form
\be\label{s2}
\frac{\eta}{r}\frac{\partial}{\partial r}r\frac{\partial}{\partial r}u_c(r)-\frac{k_BT}{er}\frac{\Delta V}{L}\frac{\partial}{\partial r}r\e(r)\frac{\partial}{\partial r}\phi(r)=0.
\ee
At this stage, we note that in the stationary regime corresponding to the constant translocation velocity, the longitudinal electric force per polymer length $F_e=-2\pi ae\sigma_p\Delta V/L$ compensates the viscous friction force $F_v=2\pi a\eta u_c'(a)$, i.e. $F_e+F_v=0$. Solving Eq.~(\ref{s2}) by accounting for this force balance relation, the Gauss' law $\phi'(a)=4\pi\ell_B\sigma_p$, and the hydrodynamic boundary conditions $u_c(d)=0$ (no-slip at the pore surface) and $u_c(a)=v$, one gets the polymer translocation velocity
\be\label{v}
v=-\mu_e\frac{\Delta V}{L}\left[\phi(d)-\phi(a)\right]
\ee
and the convective flow velocity
\be\label{uc}
u_c(r)=-\mu_e\frac{\Delta V}{L}\left[\phi(d)-\phi(r)\right],
\ee
where we introduced the electrophoretic mobility
\be
\mu_e=\frac{k_BT\e_w}{e\eta }.
\ee
Similar to the mean-field electrophoretic transport~\cite{Ghosal2007},  the translocation velocity of the polymer~(\ref{v}) is the superposition of the electrophoretic velocity of the molecule in the reference frame of the flowing liquid  $u_{e}=\mu_e\phi(a)\Delta V/L$, and the electroosmotic velocity of the charged liquid $u_{eo}=-\mu_e\phi(d)\Delta V/L$. The main difference between the present formalism and the usual mean-field transport theory is due to the correlation corrections to the electrostatic potential $\phi(r)$ in Eq.~(\ref{v}).

The ionic current thorough the pore is given by the total number of flowing ions per unit time,
\be\label{i1}
I=2\pi e\sum_iq_i\int_a^d\mathrm{d}rr\rho_i(r)u_i(r).
\ee
Substituting the ion density~(\ref{den1l}) and the velocities~(\ref{ut}) and~(\ref{uc}) into Eq.~(\ref{i1}), and expanding the result at one-loop order $O(l)$, one finds that the ionic current is given by the linear response relation $I=G\Delta V$. The total conductivity of the pore is composed of a transport flow contribution and a convective part,
\be\label{contot}
G=G_T+G_c,
\ee
with the conductive and convective components
\bea
\label{con1}
G_T&=&\frac{2\pi e}{L}\sum_i|q_i|\mu_i\int_a^d\mathrm{d}rr\rho_i(r)\\
\label{con2}
G_c&=&-\frac{2\pi e}{L}\mu_e\sum_iq_i\int_a^d\mathrm{d}rr\left\{\left[\phi_0(d)-\phi_0(r)\right]\rho_i(r)\right.\\
&&\hspace{3.5cm}+\left.\left[\phi_1(d)-\phi_1(r)\right]n_i(r)\right\}.\nonumber
\eea
In the relations~(\ref{con1}) and~(\ref{con2}), charge fluctuations are taken into account by the external potential correction $\phi_1(r)$ and the ionic self energies $\delta v^{(im,s)}(r)$ in the ion densities $n_i(r)$ and $\rho_i(r)$  (see Eqs.~(\ref{aux1}) and~(\ref{den1l})). In other words, setting these potentials to zero, one covers from Eqs.~(\ref{con1})-(\ref{con2}) the mean-field ion conductivity of the pore confining the polyelectrolyte.

\section{Results and discussion}

\subsection{Comparison with experimental pore conductivities}
\label{conex}

\begin{figure}
\includegraphics[width=1.1\linewidth]{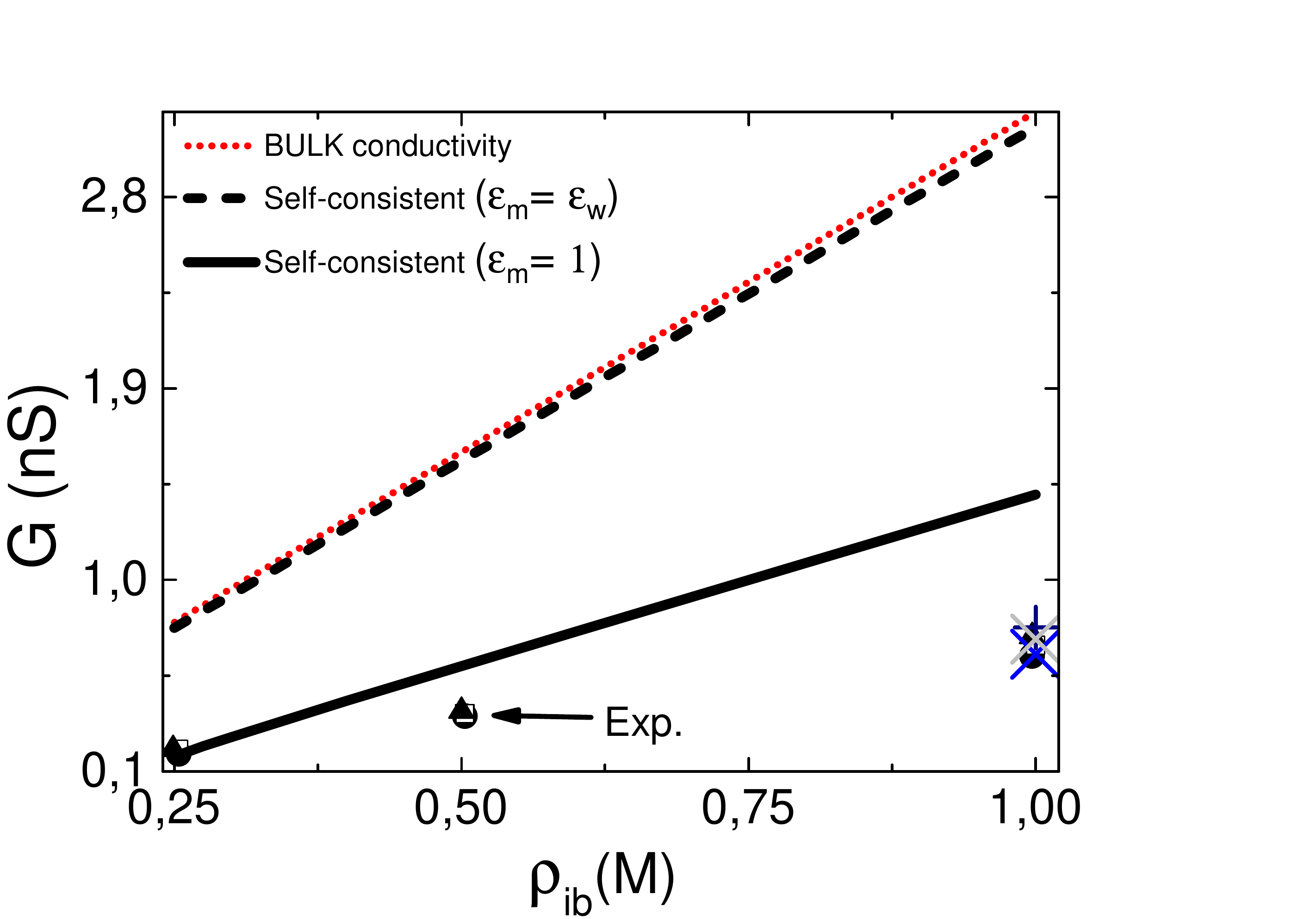}
\caption{(Color online) Conductivity of open $\alpha$-Hemolysin pores against the reservoir concentration of the KCl solution at temperature $T=281$ K. The pore modelled as an overall neutral cylinder  ($\sigma_m=0.0$ $\mathrm{C/m}^2$) has radius $d=8.5$ {\AA} and length $L=10$ nm. Experimental data : black circles, triangles, and open squares from Fig. 2 of Ref.~\cite{e3}, and additional data at $\rho_{ib}=1.0$ M from Ref.~\cite{Miles2001} (plus symbol)  and Table 1 of Ref.~\cite{Gu2000} (cross symbols).}
\label{Fig2}
\end{figure}

Within the correlation-corrected transport theory introduced above, we investigate first the mechanism behind the particularly low ionic conductivity of DNA-free $\alpha$-Hemolysin pores confining the electrolyte KCl~\cite{e3}. Because these pores exhibit a very weak cation selectivity driven by an asymmetric surface charge distribution~\cite{Noskov2004}, we model the nanopore as an overall neutral cylinder, i.e. we set $\sigma_m=0$. The radius and the length of the nanopore are taken as $d=8.5$ {\AA}~\cite{e3} and $L=10$ nm~\cite{Noskov2004}.  We display in Fig.~\ref{Fig2}  experimental conductivity data from Refs.~\cite{e3,Miles2001,Gu2000} together with three different theoretical predictions in order to investigate the effect of charge correlations.

The bulk conductivity can be derived from Eqs.~(\ref{con1})-(\ref{con2}) by neglecting the potentials $\phi(r)$ and $\delta v_{el}(r)$, which yields $G=\pi e\rho_{ib}(\mu_++\mu_-)d^2/L$. Fig.~\ref{Fig2} shows that the bulk result overestimates the experimental data by almost an order of magnitude. Furthermore, the SC prediction of Eq.~(\ref{contot}) with the dielectrically homogeneous pore approximation $\e_m=\e_w$  stays very close to the bulk conductivity curve. Thus, correlation effects solely associated with the inhomogeneity of the electrolyte play a perturbative role. However, the SC prediction of Eq.~(\ref{contot})  that takes into account the low dielectric permittivity of the membrane $\e_m=1$ agrees with the experimental data in a quantitative fashion at low ion densities and qualitatively up to about $1$ M, considerably improving on the bulk result.  The reduced pore conductivity with the lower membrane permittivity $\e_m=1\ll\e_w=80$ results from repulsive image-charge interactions between ions and the membrane. Because the radius of $\alpha$-Hemolysin pores is comparable to the Bjerrum length $d\sim\ell_B$, the particularly strong amplitude of image-charge forces lead to the dielectric exclusion of ions from the nanopore~\cite{Buyuk2014}, resulting in a pore conductivity significantly below the diffusive bulk conductivity. This indicates that the low ion permeability of DNA-free $\alpha$-Hemolysin pores stems mainly from surface polarization effects.  

\textcolor{black}{We emphasize that in Fig.~\ref{Fig2}, the deviation of the SC result from the data at large concentrations is likely to result from the dielectric continuum approximation of our solvent-implicit theory. Considering the strong confinement of $\alpha-$Hemolysin pores, the charge structure of water solvent is expected to affect their conductivity in a significant way. Indeed, within a recently developed solvent-explicit theory in nanoslits, we have shown that the solvent charge structure results in an ionic Born energy difference between the pore and the reservoir, which in turn amplifies the ionic exclusion associated with image-charge interactions~\cite{Buyuk2014II}. This suggests that the consideration of the solvent charge structure is expected to lower the ion conductivity curve of Fig.~\ref{Fig2}. However, the formidable task of integrating the solvent-explicit SC equations in cylindrical pores is beyond the scope of the present article.}

\begin{figure}
\includegraphics[width=1.1\linewidth]{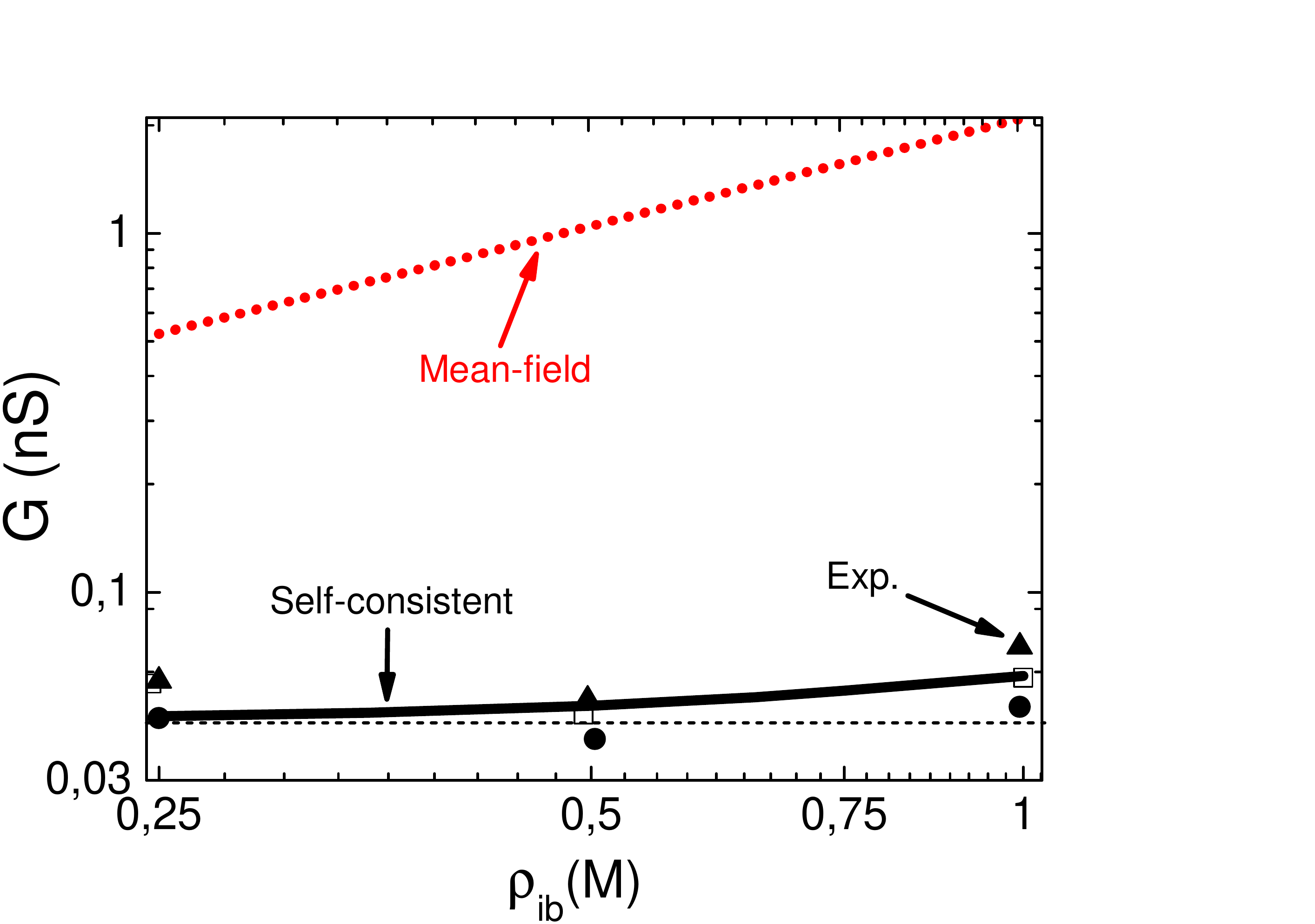}
\caption{(Color online) Conductivity of $\alpha$-Hemolysin pores blocked by an ss-DNA molecule against the reservoir concentration of the KCl solution.  The nanopore has the same characteristics as in Fig.~\ref{Fig2}. The ss-DNA molecule has radius $a=5.0$ {\AA}~\cite{e3}, surface charge $\sigma_p=0.012$ $\mathrm{e/nm}^2$, and dielectric permittivity $\e_p=50$. Experimental data (symbols) from Fig. 3 of Ref.~\cite{e3}.}
\label{Fig3}
\end{figure}

We now continue on to the transport characteristics of nanopores confining DNA molecules. We reported in Fig.~\ref{Fig3} the SC result for the salt dependence of the conductivity of an $\alpha$-Hemolysin pore blocked by a single stranded-DNA (ss-DNA) molecule, together with the experimental data from Ref.~\cite{e3}. The caption displays the dielectric permittivity and the effective smeared charge of the ss-DNA that provided the best fit with the amplitude of the experimental conductivity. First, one notes that the SC prediction can accurately reproduce the slope of the  conductivity data. Then, it is seen that unlike the conductivity of DNA-free pores linear in the ion density (see Fig.~\ref{Fig2}), the conductivity of the blocked pore varies weakly at dilute salt concentrations $\rho_{ib}\lesssim 0.5$ M but increases with salt at larger concentrations. We also reported in Fig.~\ref{Fig3} the MF conductivity obtained from Eqs.~(\ref{con1})-(\ref{con2}) by neglecting the correlation corrections, i.e. by setting $\delta v(r)=0$ and $\phi_1(r)=0$. The MF result exhibits a linear increase with ion density, which indicates that the peculiar shape of the conductivity data is a non-MF effect.

Within a phenomenological model, \textit{Bonthuis et al.} explained the non-monotonical slope of the conductivity data by the formation of neutral pairs between the DNA charges and their counterions~\cite{e3}. The present SC transport theory can bring a more physical insight into this peculiarity. For the model parameters in Fig.~\ref{Fig3}, we found that the convective part of the conductivity $G_c$ in Eq.~(\ref{con2})  is largely dominated by the transport component $G_T$ of Eq.~(\ref{con1}). In the supplemental material, by making use of a constant Donnan potential approximation,  the transport part of the conductivity is put in the closed-form 
\bea\label{conT}
G_T&\simeq&\frac{\pi e}{L}(d^2-a^2)\rho_{ib}\left\{\mu_+\frac{\Gamma^2}{\sqrt{t^2+\Gamma^2}-t}\right.\nonumber\\
&&\left.\hspace{2.5cm}+\mu_-\left[\sqrt{t^2+\Gamma^2}-t\right]\right\},
\eea
with the dielectric exclusion coefficient 
\be\label{gam}
\Gamma=\frac{2}{d^2-a^2}\int_a^d\mathrm{d}rr\;e^{-\frac{q_i^2}{2}\delta v_{el}^{(im)}(r)}
\ee
and the ratio between the pore volume density of the fixed charges and the bulk ionic density
\be\label{t}
t=\frac{\sigma_pa+\sigma_md}{\rho_{ib}(d^2-a^2)}.
\ee
In the highly confined space between the low dielectric membrane and the ss-DNA, strong image-charge interactions result in a pronounced dielectric exclusion of ions, i.e. $\Gamma\ll1$. Expanding Eq.~(\ref{conT})  up to the order $O(\Gamma^2)$, the pore conductivity follows as
\bea\label{limlaw}
G_T&\simeq&\frac{2\pi e}{L}\mu_+(\sigma_pa+\sigma_md)\nonumber\\
&&+\frac{\pi e}{2L}(\mu_++\mu_-)\frac{(d^2-a^2)^2}{\sigma_pa+\sigma_md}\Gamma^2\rho_{ib}^2.
\eea
\begin{figure}
\includegraphics[width=1.1\linewidth]{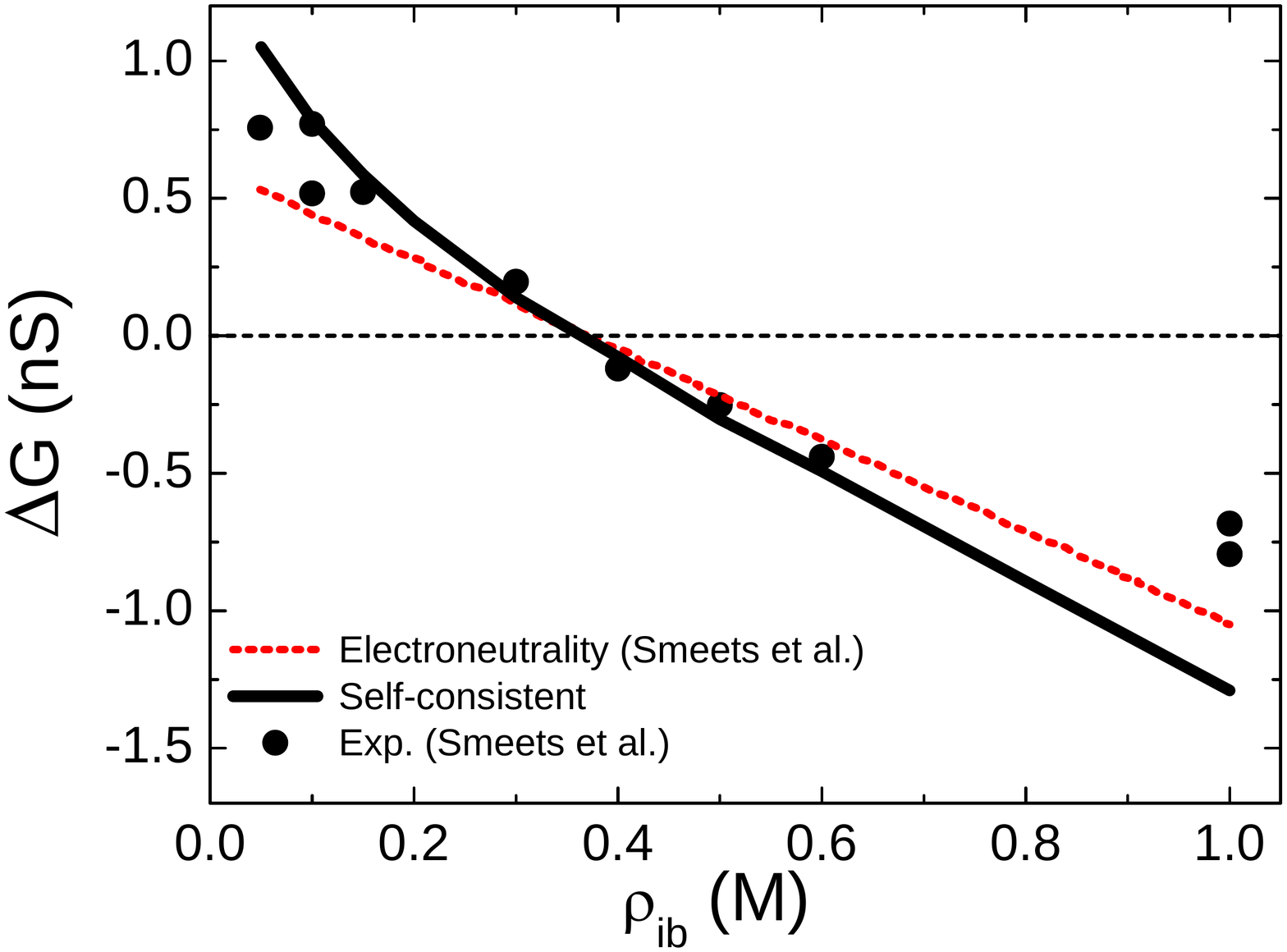}
\caption{(Color online) Conductivity change $\Delta G= G(a)-G(a\to0)$ in solid-state pores upon the penetration of a ds-DNA against the bulk concentration of the KCl  solution at temperature $T=300$ K. Experimental data as well as the value of the pore radius $d=5$ nm, length $L=34$ nm,  and surface  charge $\sigma_m=0.06$ $\mathrm{C/m}^2$  are taken from Ref.~\cite{e9}. In our SC calculation, the membrane permittivity is $\e_m=1$. The ds-DNA molecule has radius $a=1.0$ nm, surface charge $\sigma_p=0.4$ $\mathrm{e/nm}^2$, and dielectric permittivity $\e_p=50$. }
\label{Fig4}
\end{figure}
The first term of Eq.~(\ref{limlaw}) plotted in Fig.~\ref{Fig3} by the dashed horizontal curve depends solely on the ionic mobility of counterions. Thus, the ionic exclusion limit $\Gamma\to 0$ corresponds to a non-MF counterion only regime. This salt-free limit is set by the coupling between polarization forces and the electroneutrality condition of the DNA molecule. Indeed, image-charge forces repelling ions from the nanopore cannot lead to a total ionic depletion since a minimum number of counterions have to stay in the pore in order to screen the DNA charges. Thus, at low salt densities, these counterions solely contribute to the ionic current. Then, the second term of Eq.~(\ref{limlaw}) quadratic in salt density is seen to depend on the ionic mobilities of both coions and counterions. Therefore, at large ion densities, coions and counterions contribute together to the pore conductivity, which  explains the rise of the conductivity with salt concentration in Fig.~\ref{Fig3}.

Having characterized correlation effects in highly confined $\alpha$-Hemolysin pores, we now consider the role of correlations in solid-state pores of larger radius. Fig.~\ref{Fig4} displays experimental data from Fig.4(a) of Ref.~\cite{e9} for the conductivity change upon the penetration of a double stranded DNA (ds-DNA) molecule into a solid-state pore of radius $d=5$ nm. We also show our SC prediction and a linear MF result from Ref.~\cite{e9}. In the SC calculation,  the effective smeared charge of the ds-DNA  was determined as $\sigma_p=0.4$ $\mathrm{e/nm}^2$ in order to coincide with the characteristic density $\rho_{ib}\simeq0.4$ M where conductivity data changes its sign. The ds-DNA radius is assumed to be twice as large as the ss-DNA radius, that is $a=1$ nm. The conductivity change by the DNA molecule in Fig.~\ref{Fig4} was explained in Ref.~\cite{e9} by an interpolation between the low density regime $\rho_{ib}\lesssim0.4$ M where counterion attraction to DNA amplifies the ionic current ($\Delta G>0$), and the high density regime $\rho_{ib}\gtrsim0.4$ M where the DNA volume blocks the net current ($\Delta G<0$). One notes that at low ion densities, the SC theory slightly improves over the linear MF result, which is mainly due to the consideration of non-linearities in Eq.~(\ref{eqt1}) rather than charge correlations.  This indicates that in large solid-state pores confining monovalent salt, charge correlations play a perturbative role.

\subsection{Charge correlation effects on the electrophoretic motion of polyelectrolytes}
\label{corrmot}

We explore next the possibility to control the motion of polymers via charge correlation effects in synthetic pores.  The polyelectrolyte radius will be fixed to the radius of the ds-DNA as in Fig.~\ref{Fig4}.  Because the radius of solid-state pores that can contain unfolded ds-DNA molecules is usually larger than $2$ nm~\cite{Wanunu2012}, the pore radius will be taken as $d=3$ nm, unless stated otherwise. In the previous section, we found that  image-charge forces are perturbative in large pores of radius $d\gg \ell_B$. Thus, to simplify the numerical task and the physical picture, we will neglect the dielectric jumps and set $\e_m=\e_p=\e_w$. We will also assume that the pore is neutral $\sigma_m=0$ and the liquid is at ambient temperature $T=300$ K. With these model parameters, we characterize first the effect of charge correlations on the electrophoretic motion of polyelectrolytes in the simplest case of an asymmetric electrolyte. Then,  we investigate the possibility to tune the translocation velocity of polyelectrolytes by changing the counterion concentrations in electrolyte mixtures.

\subsubsection{Correlation effects in asymmetric electrolytes}
\label{transcoun}

\begin{figure}
\includegraphics[width=1.1\linewidth]{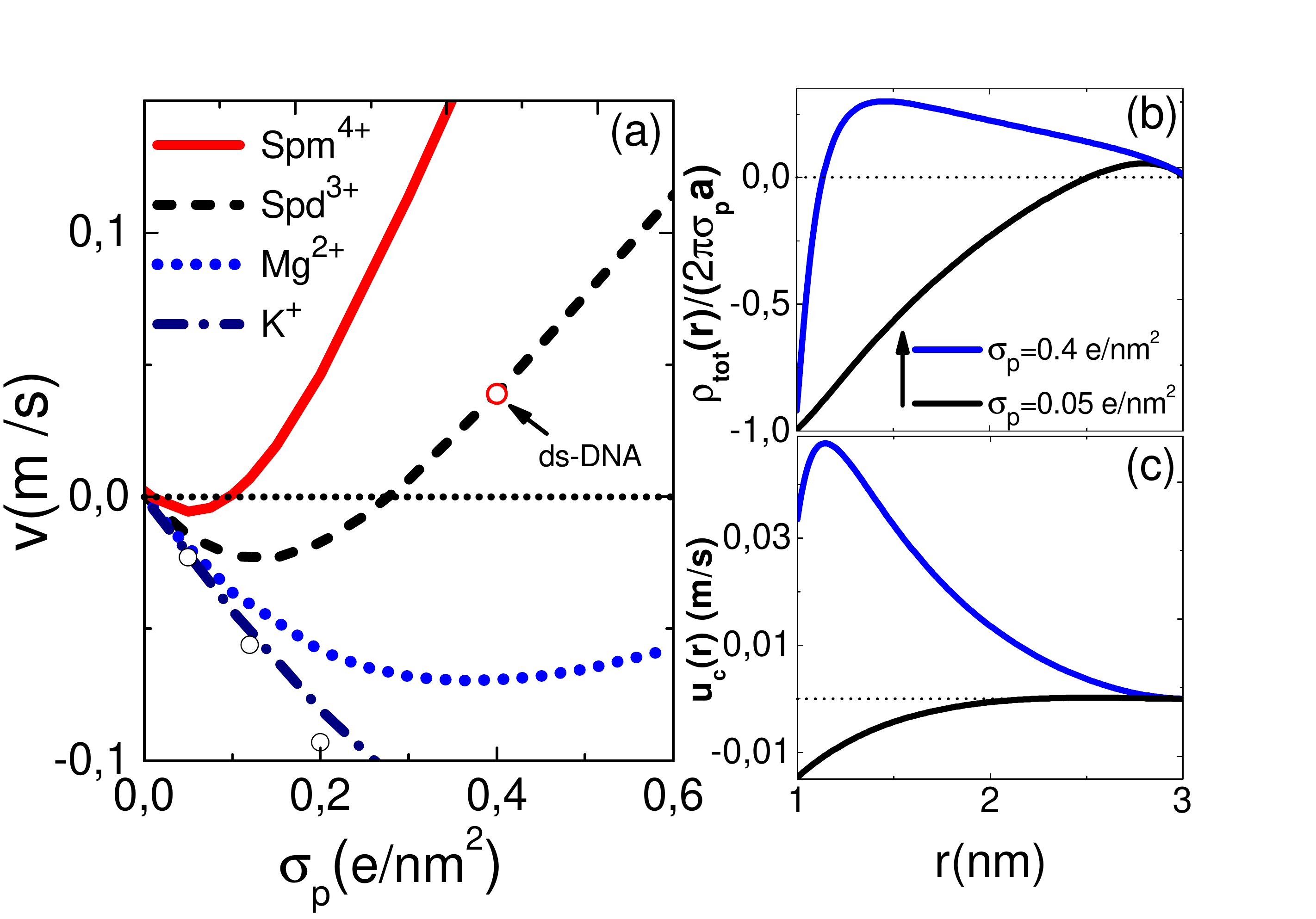}
\caption{(Color online)  (a) Velocity of the polyelectrolyte against its surface charge in an asymmetric electrolyte I$\mbox{Cl}_m$. The counterion species $\mbox{I}^{m+}$  specified in the legend has reservoir density $\rho_{mb}=0.01$ M. Open black circles mark the MF prediction of Eq.~(\ref{vellimlaw}). (b) Cumulative charge density of the $\mbox{Spd}^{3+}$ liquid rescaled with the polymer charge and (c) solvent velocities. The potential gradient is $\Delta V=120$ mV. The neutral pore has radius $d=3$ nm, length $L=34$ nm, and permittivity $\e_m=\e_w$. The confined polyelectrolyte has radius $a=1.0$ nm and permittivity $\e_p=\e_w$.}
\label{Fig5}
\end{figure}

We illustrate in Fig.~\ref{Fig5}(a) the translocation velocity of a polyelectrolyte against its surface charge in the presence of an asymmetric electrolyte with composition I$\mbox{Cl}_m$. The counterion species $\mbox{I}^{m+}$ of reservoir concentration $\rho_{mb}=0.01$ M is specified in the legend. One notes that with monovalent $\mbox{K}^+$ counterions, the translocation velocity is negative for all polymer charges, i.e. the polymer moves oppositely to the applied field. By evaluating the translocation velocity~(\ref{v}) with the electrostatic potential Eq. (3) of the supplemental material, taking the MF dilute salt limit $\Gamma\to0$ and $\rho_{ib}\to0$, and Taylor-expanding the result in terms of the polymer surface charge $\sigma_p$, the translocation velocity becomes
\be\label{vellimlaw}
v=-2\pi\ell_Ba\sigma_p\left[\frac{2d^2}{d^2-a^2}\ln\left(\frac{d}{a}\right)-1\right]\mu_e\frac{\Delta V}{L}.
\ee
Eq.~(\ref{vellimlaw}) reported in Fig.~\ref{Fig5} by open circles is seen to closely follow the SC result. Thus, with monovalent electrolytes, the polymer motion is qualitatively driven by MF electrophoretic transport and charge correlations weakly affect the polymer translocation. Then, with multivalent counterions, the amplitude of the velocity is seen in Fig.~\ref{Fig5}(a) to increase also according to the linear law~(\ref{vellimlaw}) for weak polymer charges, but the slope of the curves vanishes at a characteristic polymer charge and the velocity starts decreasing beyond this value. With $\mbox{Mg}^{2+}$ ions, the polymer velocity 
stays always negative in the physical charge regime considered in Fig.~\ref{Fig5}(a). With spermidine or spermine molecules of higher valency,  the translocation velocity vanishes  ($v=0$) respectively  at the polymer charges $\sigma_p\simeq 0.3$ $\mbox{e/nm}^2$ and $\sigma_p\simeq 0.1$ $\mbox{e/nm}^2$. Polyelectrolytes with stronger charges translate in the direction of the external field ($v>0$), a peculiarity that cannot be explained by the MF prediction of Eq.~(\ref{vellimlaw}).

In order to elucidate the underlying mechanism behind the reversal of the polymer motion, we show in Fig.~\ref{Fig5}(b) and (c) the cumulative charge density of the $\mbox{Spd}^{3+}$ liquid and its electroosmotic velocity  Eq.~(\ref{uc}). The cumulative charge density including the polyelectrolyte is defined as the integrated charge
\be\label{cum}
\rho_{tot}(r)=2\pi\int_a^r\mathrm{d}r'r'\left[\rho_c(r')+\sigma_p(r')\right].
\ee
At the surface charge $\sigma_p=0.05$ $\mathrm{e/nm}^2$ where the polymer translocates oppositely to the applied field  ($v<0$ in Fig.~\ref{Fig5}(a)), the counterion screening of the polymer results in an overall negative cumulative charge and convective velocity. By increasing the polymer charge to the characteristic value of the ds-DNA $\sigma_p=0.4$ $\mathrm{e/nm}^2$, in the close neighborhood of the polymer surface $r\simeq 1.1$ nm, the cumulative charge becomes positive, which inverses the sign of the convective velocity of the liquid and the DNA molecule. Thus, the inversion of the translocation velocity results from the reversal of the polymer charge, i.e. the effect is driven by electrostatic correlations between multivalent counterions bound to the polyelectrolyte. The overcompensation of the macromolecular charge by strongly correlated multivalent counterions has been observed by previous MD simulations~\cite{Messina2001,Messina2002} and nanofluidic experiments~\cite{Heyden2006}.  The phenomenon observed in Fig.~\ref{Fig5}(a) presents itself as a useful mechanism to minimize the velocity of translocating polymers. However, the charge of a polyelectrolyte cannot be easily tuned in translocation experiments. Thus, we will next explore the possibility to control this effect via counterion concentrations in electrolyte mixtures.

\begin{figure}
\includegraphics[width=1.2\linewidth]{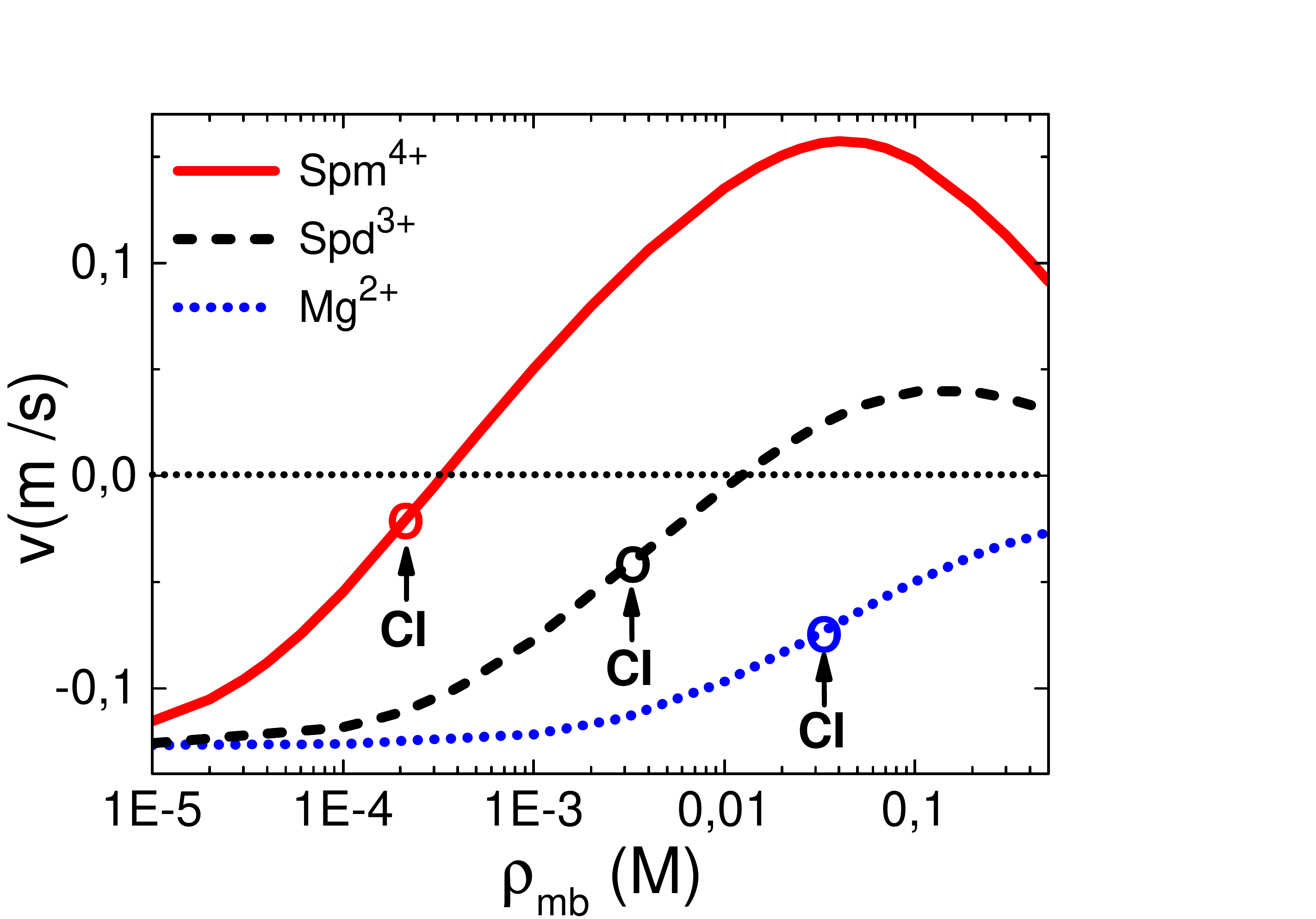}
\caption{(Color online)  Polymer translocation velocity against the density of the multivalent counterion species $\mbox{I}^{m+}$  (see the legend) in the electrolyte mixture KCl+I$\mbox{Cl}_m$. $\mbox{K}^+$ density is fixed at $\rho_{+b}=0.1$ M. The confined ds-DNA has surface charge $\sigma_p=0.4$ $\mathrm{e/nm}^2$. The remaining parameters are the same as in Fig.~\ref{Fig5}. Circles mark the charge inversion (CI) point of the DNA molecule.}
\label{Fig6}
\end{figure}

\subsubsection{Tunning translocation velocity in electrolyte mixtures via counterion densities}
\label{transsalt}

We consider an electrolyte mixtures  of composition KCl+I$\mbox{Cl}_m$ containing an arbitrary type of multivalent counterion $\mbox{I}^{m+}$.  In this part, the polymer charge will be set to the characteristic value of ds-DNA $\sigma_p=0.4$ $\mbox{e/nm}^2$, unless stated otherwise. We display in Fig.~\ref{Fig6} the DNA translocation velocity against the reservoir density of three different types of multivalent counterions in the solution (see the legend). The bulk $\mbox{K}^+$ density is fixed at $\rho_{+b}=0.1$ M. For each type of multivalent counterion, the increase of the reservoir density first results in the inversion of the DNA charge at $\rho_{2+}\simeq3\times10^{-2}$ M, $\rho_{3+}\simeq3\times10^{-3}$ M, and $\rho_{4+}\simeq2\times10^{-4}$ M. For electrolytes with spermidine and spermine molecules, this is followed by the blockage of the translocation ($v=0$) at the critical densities $\rho_{3+}\simeq1.2\times10^{-2}$ M and $\rho_{4+}\simeq3.5\times10^{-4}$ M. Increasing further the  counterion density, the translocation velocity that becomes positive reaches a peak and starts decreasing. With divalent magnesium ions, the polymer velocity stays negative for all $\mbox{Mg}^{2+}$ concentrations, i.e. the charge inversion is not strong enough to reverse the motion of DNA. This conclusion agrees with MD simulations of electrophoretic DNA motion in charged liquids~\cite{Luan2009}.

\begin{figure}
\includegraphics[width=1.1\linewidth]{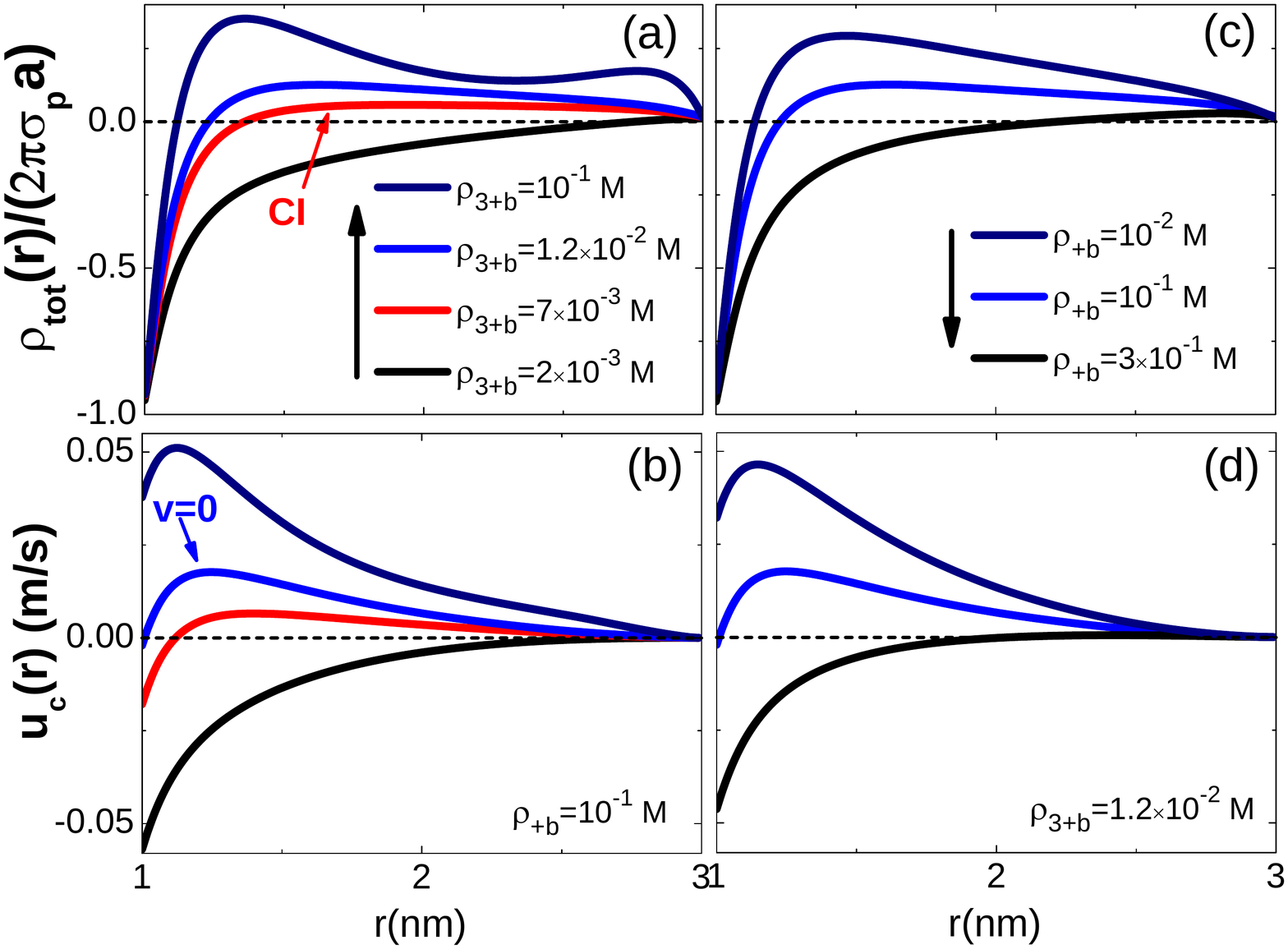}
\caption{(Color online) Cumulative charge densities rescaled with the bare DNA charge (top plots) and solvent velocities (bottom plots) at fixed $\mbox{K}^+$ and varying $\mbox{Spd}^{3+}$ concentration in (a) and (b), and fixed $\mbox{Spd}^{3+}$ and varying  $\mbox{K}^+$ concentration in (c) and (d). In each column, the same colour in the top and bottom plots corresponds to a given bulk counterion  concentration displayed in the legend. The remaining parameters are the same as in Fig.~\ref{Fig6}.}
\label{Fig7}
\end{figure}

To illustrate the physical picture behind these observations, we report in Fig.~\ref{Fig7}(a) and (b) the cumulative charge density and the electroosmotic velocity of the fluid for various $\mbox{Spd}^{3+}$ concentrations. It is seen that with an increase of the $\mbox{Spd}^{3+}$ density from $\rho_{3+b}=2\times10^{-3}$ M to $7\times10^{-3}$ M, the cumulative charge density changes its sign and becomes positive at  $r\simeq 1.3$ nm. As a result of this charge inversion, the liquid flows in the direction of the applied field ($u_c(r)>0$) in the region $r\gtrsim 1.3$ nm. However, at the corresponding  $\mbox{Spd}^{3+}$ density, the hydrodynamic drag is not strong enough to reverse the motion of DNA, and the DNA molecule as well as the fluid carried by the latter at $r\lesssim 1.3$ nm continue to move oppositely to the field. By increasing further the  $\mbox{Spd}^{3+}$ density to the characteristic value $\rho_{3+b}=1.2\times10^{-2}$ M, the cumulative charge density becomes positive enough for the hydrodynamic drag to compensate exactly the electrostatic coupling between DNA and the applied field. As a result, the translocation velocity of DNA vanishes. For the higher reservoir density $\rho_{3+b}=10^{-1}$ M, the DNA and the surrounding fluid move in the direction of the applied field. In Fig.~\ref{Fig7}(a), one also notes that the cumulative charge density develops a well between the two density peaks located at $r\simeq1.3$ nm and $r\simeq2.8$ nm.  The corresponding local decrease in the cumulative charge density results from the enhanced concentration of $\mbox{Cl}^-$ coions attracted to the charge inverted DNA molecule. This $\mbox{Cl}^-$ layer attracts in turn $\mbox{K}^+$ ions, leading to the second peak close to the pore wall. Indeed, we found that with a further increase of the $\mbox{Spd}^{3+}$ concentration, the intensification of the $\mbox{Cl}^-$ attraction starts lowering the cumulative charge and the convective velocity. This explains the reduction of the positive translocation velocity at large $\mbox{Spd}^{3+}$ concentrations $\rho_{3+b}\gtrsim 0.1$ M in Fig.~\ref{Fig6}.

\begin{figure}
\includegraphics[width=1.1\linewidth]{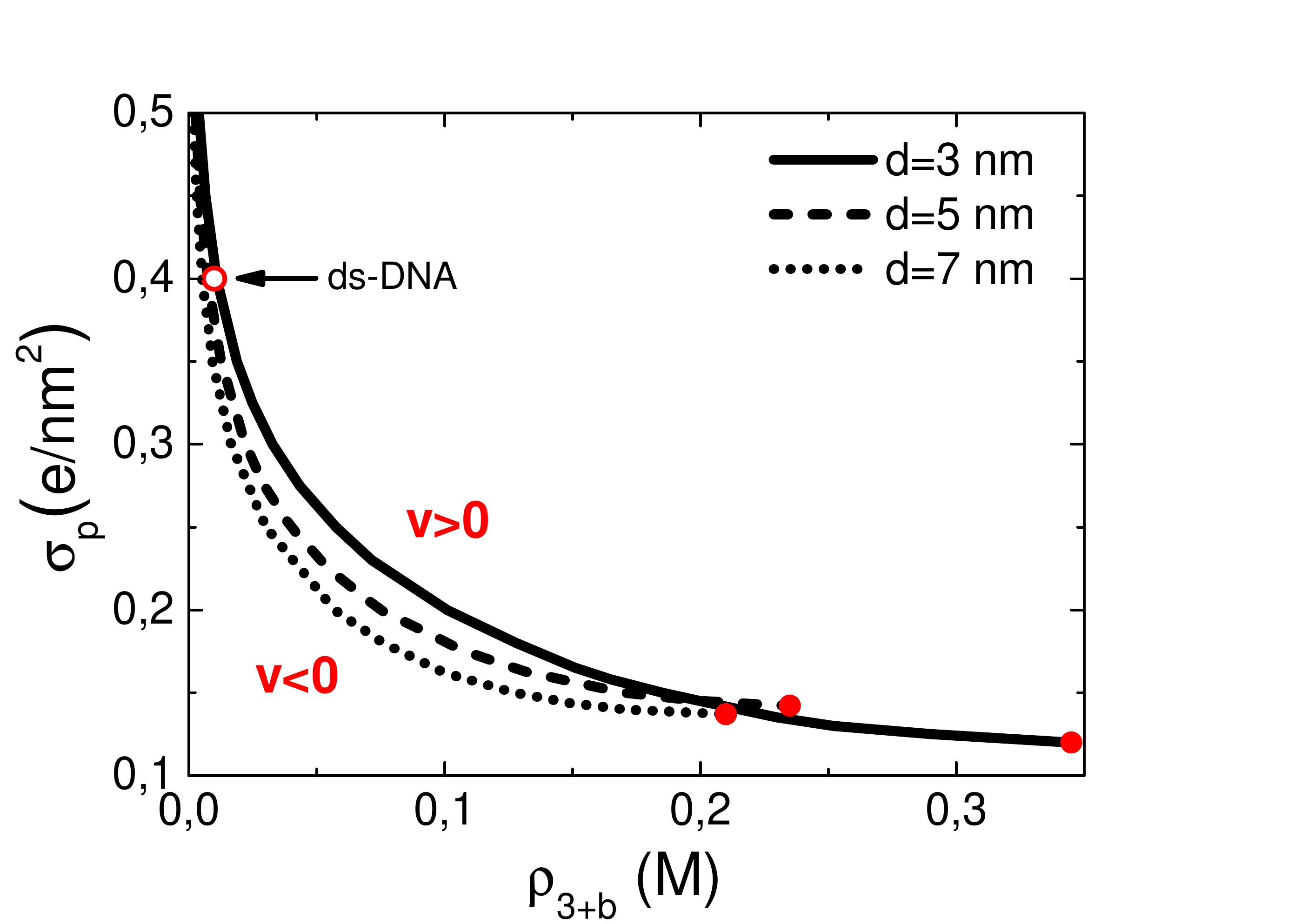}
\caption{(Color online) Characteristic polymer charge $\sigma_p^*$ versus $\mbox{Spd}^{3+}$ concentration  $\rho_{3+b}^*$  curves separating the charge density regime with positive translation velocity $v>0$ (above the curves) and negative translocation velocity $v<0$ (below the curves). The corresponding pore radii are given in the legend. $\mbox{K}^+$ density is fixed at $\rho_{+b}=0.1$ M. The remaining parameters are the same as in Fig.~\ref{Fig6}.}
\label{Fig8}
\end{figure}

These results indicate that the modification of the multivalent counterion density is an efficient way to minimize the translocation velocity of polyelectrolytes. In order to guide future translocation experiments aiming at exploring this effect, we illustrate in Fig.~\ref{Fig8}  the characteristic polymer charge $\sigma_p^*$ versus $\mbox{Spd}^{3+}$ density $\rho_{3+b}^*$ curves separating the parameter domains with positive and negative translocation velocity. First, the phase diagram indicates that weaker is the polymer charge, higher is the $\mbox{Spd}^{3+}$ density required to cancel the translocation velocity, i.e. $\sigma_p^*\downarrow\rho_{3+b}^*\uparrow$. However, one also sees that with decreasing polymer charge density, the characteristic curves end at a \textit{critical point} located at $\sigma_p^*\simeq 0.1-0.15$ $\mbox{e/nm}^2$ and $\rho_{3+b}^*\simeq 0.2-0.35$ M where the inversion of the polyelectrolyte motion ceases to exist. At lower polymer charges, the translocation velocity stays always negative regardless of the $\mbox{Spd}^{3+}$ concentration in the reservoir. Indeed, we found that below this critical polymer charge, the peak in the velocity curves of Fig.~\ref{Fig6} emerges before the velocity becomes positive. This observation fixes a lower polymer charge boundary for the inversion of the polymer motion to occur. Nevertheless, for the physiological $\mbox{K}^+$ density $\rho_{+b}=0.1$ M in Fig.~\ref{Fig8}, the ds-DNA charge is shown to be located well above this lower boundary. Finally, the phase diagram in Fig.~\ref{Fig8} indicates that in smaller pores, the main effect of confinement is a shift of the critical $\mbox{Spd}^{3+}$ densities to larger values, that is $d\downarrow\rho_{3+b}^*\uparrow$ at fixed polymer charge $\sigma_p^*$. However, one also notes that the effect of the pore radius on the critical curves is only moderate.

\begin{figure}
\includegraphics[width=1.1\linewidth]{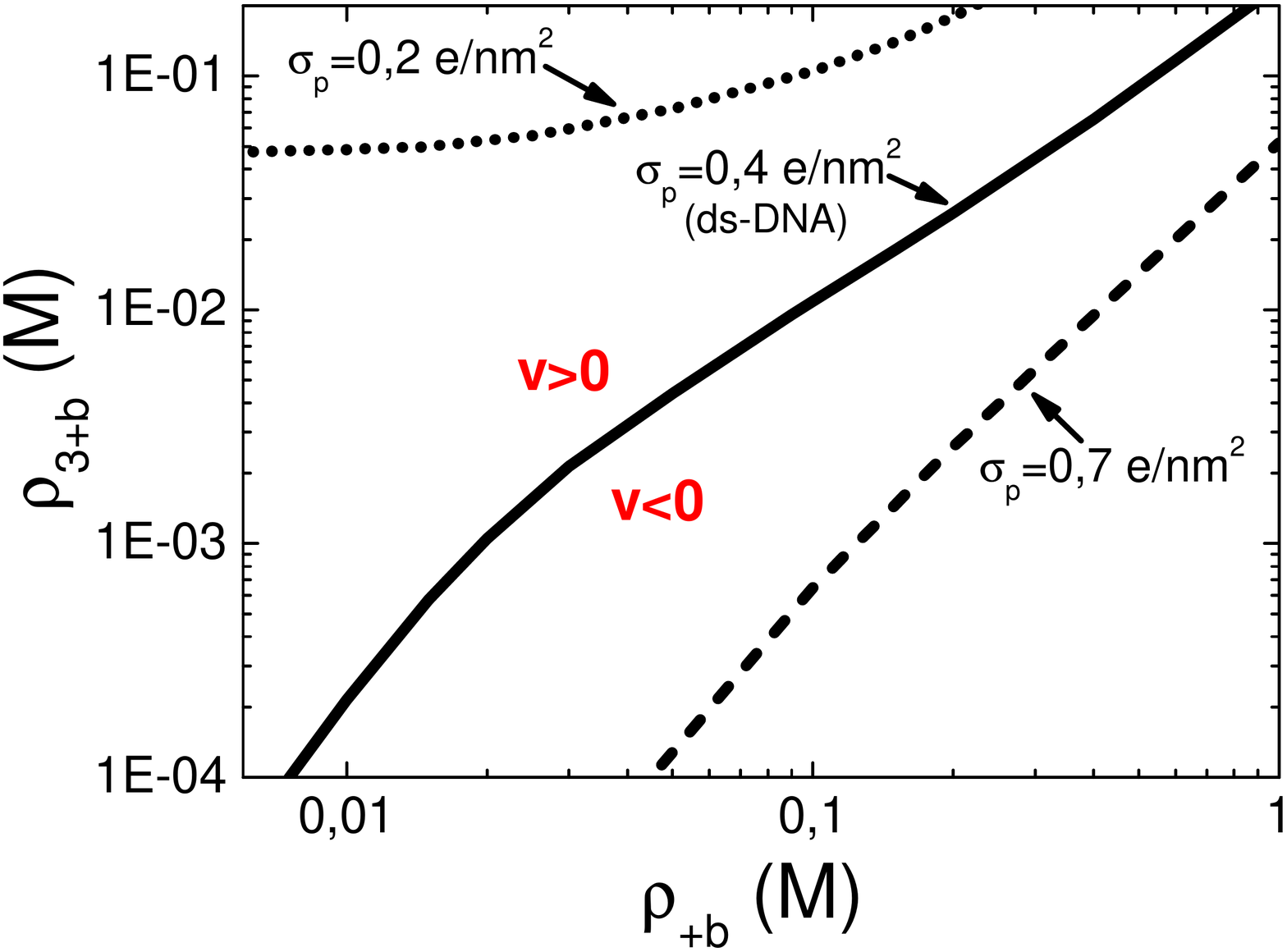}
\caption{(Color online) Characteristic $\mbox{Spd}^{3+}$ concentration $\rho_{3+b}^*$ versus $\mbox{K}^{+}$  concentration $\rho_{+b}^*$ curves separating the charge density regime with positive translation velocity $v>0$ (above the curves) and negative translocation $v<0$ velocity (below the curves). The corresponding polymer charge $\sigma_p$ is specified for each curve. The remaining parameters are the same as in Fig.~\ref{Fig6}.}
\label{Fig9}
\end{figure}

An additional parameter that can be tuned in translocation experiments is the $\mbox{K}^+$ concentration in the reservoir. We characterized the effect of $\mbox{K}^+$ ions on the polyelectrolyte motion in Fig.~\ref{Fig9}. The figure displays the characteristic $\mbox{K}^+$ density $\rho_{+b}^*$ versus $\mbox{Spd}^{3+}$ density $\rho_{3+b}^*$ curves splitting the domains with positive and negative translocation velocity. The phase diagram indicates that at fixed $\mbox{Spd}^{3+}$ density and starting at any point above the curves, the increase of $\mbox{K}^+$ density switches the translocation velocity from positive to negative. Then, with increasing polymer charge, the critical lines are seen to move towards higher $\mbox{K}^{+}$ concentrations, i.e. $\sigma_p\uparrow\rho_{+b}^*\uparrow$ at fixed $\mbox{Spd}^{3+}$ concentration in the reservoir. This means that stronger is the polymer charge, higher is the amount of $\mbox{K}^{+}$ ions needed to stop the translocation of the molecule. To better understand the physical effect of $\mbox{K}^+$ ions on the translocation velocity,  in Fig.~\ref{Fig7}(c), we show that the increase of the  $\mbox{K}^+$ concentration from $\rho_{+b}=0.01$ M to $0.1$ M reduces the overscreening of the DNA charge. The effect is totally suppressed at the higher concentration $\rho_{+b}=0.3$ M.  \textcolor{black}{We note that monovalent salt induced suppression of the macromolecular charge reversal in asymmetric electrolyte mixtures had been observed and quantitatively characterized in previous modified PB theories~\cite{Pianegonda2005,Diehl2006,dosSantos2010}.} In Fig.~\ref{Fig7}(d), one sees that this \textcolor{black}{effect} is accompanied with the blocking of the translocation process and then the switching of the liquid and DNA velocities from positive to negative. This observation suggests the alteration of the  $\mbox{K}^+$ concentration as an alternative way to minimize the velocity of DNA in translocation experiments.

\section{Conclusions}

We developed a correlation-corrected theory in order to characterize charge and polymer transport properties of membrane nanopores in physiological conditions where the PB theory is invalid. First, the notably reduced conductivities of  $\alpha$-Hemolysin pores were found to result from the presence of polarization charges induced by the dielectric mismatch between the electrolyte and the membrane. Then, we showed that in the presence of an ss-DNA in the pore, the same polarization effects combined with the electroneutrality of the DNA drive the system into a counterion-only regime, setting the lower boundary of the ionic current flowing through the blocked pore. Finally, we characterized charge correlation effects on the electrophoretic translocation of polyelectrolytes. We found that the presence of multivalent counterions in the solution results in the reversal of the polymer charge and the electroosmotic velocity of the liquid. With $\mbox{Spd}^{3+}$ and $\mbox{Spm}^{4+}$ ions, the reversal of the fluid velocity is strong enough to invert the electrophoretic motion of the polyelectrolyte. By adequately tuning the monovalent or multivalent counterion density in the reservoir, this mechanism can be efficiently used to minimize the DNA velocity in translocation experiments. 

The present model can be elaborated further by considering the finiteness of the nanopore length~\cite{Levin2006}, or the helicoidal geometry and the discrete charge structure of DNA molecules~\cite{Golestanian2004,Sung2013}. \textcolor{black}{We should also note that our theory cannot account for ionic pair formation observed in previous MC simulations and modified PB approaches~\cite{Pianegonda2005,Diehl2006,dosSantos2010}. The formation of such pairs between monovalent coions and trivalent/quadrivalent molecules should lower the charge of the latter and consequently weaken the inversion of the polymer mobility. Thus, the present theory that misses ionic cluster formation may underestimate the multivalent charge concentrations and the polymer charge densities where the inversion of the polymer velocity takes place. The consideration of the cluster formation in future works will probably require the inclusion of second order cumulant corrections to the variational grand potential of the system. Then, our formalism lacks as well the charge structure of solvent molecules. This dielectric continuum approximation may indeed be responsible for the overestimation of the pore conductivity at large ion densities in Fig.~\ref{Fig2}. Thus, future works should consider the solvent charge structure in modelling ionic conductivity through nanoscale pores~\cite{Buyuk2014II}.}  An additional detail that deserves consideration is the flexibility of polyelectrolytes~\cite{Duncan1999}. Although the rigidly cylindrical polyelectrolyte model allowed us to consider surface polarization effects, the considerably low value of the effective ss-DNA charge in Fig.~\ref{Fig3} is likely to result from charge discreteness and/or DNA configuration effects neglected herein. We emphasize that despite these model simplifications, the present theory can already capture different characteristics of membrane nanopores. Moreover, by including charge correlations, the formalism bridges a gap between MF transport theories unable to consider multivalent charges or surface polarization effects~\cite{Ghosal2006,Ghosal2007,Keijan2009} and MD simulations with considerable complexity~\cite{Luan2009}. Future translocation experiments with multivalent ions will be certainly needed to ascertain our physical conclusions.
\\

{\bf Acknowledgement.}  We thank Ralf Blossey for a detailed reading of our manuscript and his precious comments. S.B. gratefully acknowledges support under the ANR blanc grant ``Fluctuations in Structured Coulomb Fluids''. T. A-N. has been supported in part by the Academy of Finland through its CoE program COMP grant no. 251748.

\smallskip
\appendix
\section{Derivation of the pore conductivity within Donnan approximation}
\label{ap1}

In this appendix, we derive a close-form expression for the conductivity of DNA-blocked pores confining the monovalent symmetric electrolyte KCl.  In strongly confined $\alpha$-Hemolysin pores, correlation effects resulting from the inhomogeneous screening of the electrolyte is dominated by pronounced image-charge forces. Thus, our first approximation consists in neglecting the solvation potentials $\phi_1(r)$ and $\delta v^{(s)}(r)$ in Eq.~(\ref{con1}).  As a further approximation, we replace the Boltzmann factor containing the image charge potential by its average over the cross section of the nanopore, and cast the conductivity in the form
\be
\label{conap1}
G_T\simeq\frac{2\pi}{L}e\rho_{ib}\Gamma\int_a^d\mathrm{d}rr\left[\mu_+e^{-\phi_0(r)}+\mu_-e^{\phi_0(r)}\right],
\ee
where we introduced the partition coefficient taking into account exclusively the dielectric exclusion of ions
\be\label{conap2}
\Gamma=\frac{2}{d^2-a^2}\int_a^d\mathrm{d}rr\;e^{-\frac{q_i^2}{2}\delta v_{el}^{(im)}(r)}.
\ee
To progress further, we will compute the external potential in Eq.~(\ref{conap1}) within a Donnan potential approximation.  To this aim, we split the potential into a constant Donnan potential and a non-uniform correction,
\be
\label{don1}
\phi_0(r)=\phi_D+\delta\phi(r).
\ee
Injecting the ansatz~(\ref{don1}) into the generalized PB equation~(\ref{eqt1}),  linearizing the latter in the potential $\delta \phi(r)$, and replacing again the Boltzmann factor including the image charge potential by its average in Eq.~(\ref{conap2}), one obtains 
\bea\label{conap3}
&&\frac{1}{r}\partial_rr\partial_r\delta\phi(r)-8\pi\ell_B\rho_{ib}\Gamma\left[\sinh(\phi_D)+\cosh(\phi_D)\delta\phi(r)\right]\nonumber\\
&&=-4\pi\ell_B\sigma(r).
\eea
Neglecting first the inhomogeneous part of the potential in  Eq.~(\ref{conap3}), i.e. setting $\delta \phi(r)=0$ and integrating the remaining terms over the cross section of the pore, the Donnan potential follows as
\be\label{don2}
\phi_D=\ln\left(-\frac{t}{\Gamma}+\sqrt{\frac{t^2}{\Gamma^2}+1}\right),
\ee
with the auxiliary parameter
\be\label{t}
t=\frac{\sigma_pa+\sigma_md}{\rho_{ib}(d^2-a^2)}.
\ee
Solving now the differential equation~(\ref{conap3}) with the electrostatic boundary conditions $\delta\phi'(a)=4\pi\ell_B\sigma_p$ and $\delta\phi'(d)=-4\pi\ell_B\sigma_m$, the inhomogeneous part of the potential follows in the form
\be\label{conap4}
\delta\phi(r)=C_1I_0(\mu r)+C_2K_0(\mu r)-\tanh(\phi_D),
\ee
with the integration constants
\bea
C_1&=&\frac{4\pi\ell_B}{\mu}\frac{\sigma_pK_1(\mu d)+\sigma_mK_1(\mu a)}{K_1(\mu d)I_1(\mu a)-I_1(\mu d)K_1(\mu a)}\\
C_2&=&\frac{4\pi\ell_B}{\mu}\frac{\sigma_pI_1(\mu d)+\sigma_mI_1(\mu a)}{K_1(\mu d)I_1(\mu a)-I_1(\mu d)K_1(\mu a)},
\eea
and the auxiliary parameter $\mu^2=8\pi\ell_B\rho_{ib}\Gamma\cosh(\phi_D)$. Substituting into Eq.~(\ref{conap1}) the potential~(\ref{don1}) with the components given by Eqs.~(\ref{don2}) and~(\ref{conap4}), linearizing  the result in the potential correction $\delta\phi(r)$, and carrying out the integral over the pore, one obtains after some algebra the result~(\ref{conT}) of the main text.

\end{document}